# First Single-Carrier Transmission at Net Data Rates of 1.6 Tb/s over 9075 km and 2.4 Tb/s over 1210 km Using 300 GBd Dual-Polarization Signals and Probabilistic Constellation Shaping

Haïk Mardoyan[(1)*], Daniel Drayss[(2,3)*], Sylvain Almonacil[(1)*], Dengyang Fang[(2)*], Alban Sherifaj[(2)], Amirhossein Ghazisaeidi[(1)], Mohamed Kelany[(2)], Carina Castineiras Carrero[(1)], Christian Koos[(2,3)], Jérémie Renaudier[(1)]

[(1)] Nokia Bell Labs, Optical Transmission Dpt., Massy, France, haik.mardoyan@nokia-bell-labs.com
[(2)] Institute of Photonics and Quantum Electronics (IQP), Karlsruhe Institute of Technology, Karlsruhe, Germany, daniel.drayss@kit.edu
[(3)] Institute of Microstructure Technology (IMT), Karlsruhe Institute of Technology, Eggenstein-Leopold-Shafen, Germany
[*] Contributed equally

**Abstract** *We report long-haul transmissions of single-carrier 300 GBd dual-polarization signals with optical arbitrary waveform generation and measurement. We demonstrate net 1.6 Tb/s over 9075 km with PCS-16QAM and 2.4 Tb/s over 1210 km with PCS-36QAM. ©2024 The Author(s)*

**Introduction**

To answer the sustained traffic growth in optical networks, the main trend in the telecom industry has been the transition to higher spectral efficiency and higher symbol rates through integrated optics to reduce the overall cost per bit [1]. Recently, 800 G coherent modules have been made available on the market to support 800GE transport over long-haul networks, enabled by higher symbol rates, capacity-approaching solutions based on probabilistic constellation shaping (PCS), and optimized forward error correction (FEC) codes. To further sustain that trend, increasing the aggregate per-carrier information rate up to 1.6 Tb/s and beyond is necessary, yet requires leveraging high symbol rate interfaces above 200 GBd for long haul networks. While advanced high-speed coherent systems have been mainly relying on Silicon-Germanium (SiGe) and Complementary Metal-Oxide Semiconductor (CMOS) technologies for digital/analog conversion [2], increasing their bandwidth above 70 GHz without sacrificing on the resolution and the output swing is challenging. As a result, both electrical and optical interleaving techniques have been recently investigated to achieve single-carrier signaling at symbol rates above 200 GBd. Figure 1 shows a summary of the most recent experiments [3]-[10] demonstrating ≥ 1.6 Tb/s high data rates per carrier. With electrical interleaving only, despite major improvements brought by the proposed techniques, the reach has been mainly limited to 240 km due to the limited signal quality at symbol rates up to 260 GBd. At the same time, optical spectrum slicing has been rekindled through the demonstration of optical arbitrary waveform generation (OAWG) and measurements (OAWM) [11] producing very good signal integrity far above 200 GBd. However, long-haul transmission of optically sliced signals at around 300 GBd and above has never been investigated so far.

In this paper, we demonstrate the transmission of net data rates of 1.6 Tb/s and 2.4 Tb/s over distances of 9075 km and 1210 km, respectively, based on 300 GBd dual polarization PCS-QAM single-carrier signals. The channel under test is generated with 2-slice OAWG and received through 2-slice OAWM. It is inserted into a wavelength-division-multiplexed (WDM) comb with channel spacing of 306.25 GHz and transmitted over a recirculating loop made of ultra-low-loss large effective area EX3000 fibers separated by Erbium-doped fiber amplifiers (EDFAs). With a 10x improvement in transmission distance for aggregated capacity per carrier of 1.6 Tb/s and 2.4 Tb/s, we show the potential benefits of optical multiplexing techniques for future generations of coherent transmission systems.

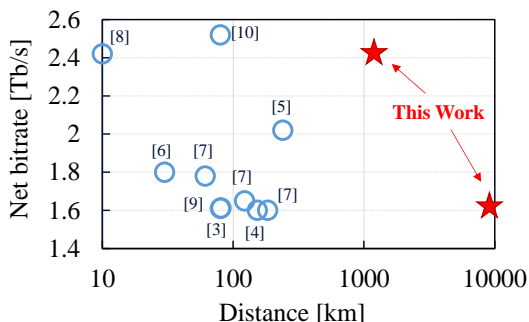

**Fig. 1:** Reported bitrates above 1.6 Tb/s versus transmission distance [3]-[10].

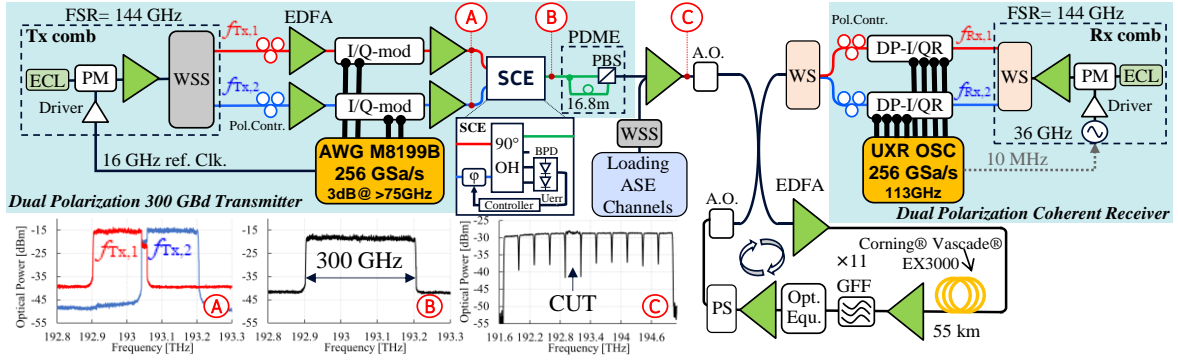

**Fig. 2:** Experimental setup. PDME: polarization division multiplexing emulator; ECL: External-cavity laser; PM: phase modulator; EDFA: erbium-doped fiber amplifier; WSS: wavelength-selective switch; WS: Wave Shaper; PS: Polarization Scrambler; AO: Acousto-Optics switch; GFF: Gain Flattening Filter; I/Q modulator (I/Q-mod); SCE: signal-combining element; OH: 90° optical hybrid; BPD: balanced photodetector; φ: phase shifter; PBS: polarization beam splitter; DP-IQR: dual-polarization IQ receiver; OSC: oscilloscope. Inset Ⓐ: Spectrum of individual slices. Inset Ⓑ: Spectrum of the 300 GBd signal. Inset Ⓒ: WDM comb spectrum at the transmitter output, comprising the channel under test (CUT) and 10 ASE dummy channels.

**Experimental setup**

The experimental setup for the 300 GBd transmission experiment is shown in Fig. 2. It comprises a dual polarization transmitter based on OAWG, a recirculating fiber loop, and a dual polarization receiver based on OAWM. To generate the dual-polarization 300 GBd signal, we first generate a transmitter frequency comb (Tx comb) by modulating a continuous-wave (CW) laser tone at 193.044 THz with a phase modulator (PM) driven by a 16 GHz RF sinusoidal signal, derived from the reference clock output of the M8199B AWG. The Tx comb is amplified by an EDFA. Two optical tones at $f_{Tx,1}$ = 192.980 THz and $f_{Tx,2}$ = 193.124 THz, spaced by a free spectral range (FSR) of 144 GHz (9×16 GHz), are separated by a wavelength selective switch (WSS). The two optical tones are modulated by two independent coherent driver modulators. The corresponding I/Q driving signals are defined to generate two spectral slices whose combination produces the single-carrier 300 GBd signal, as shown in the optical spectra in insets A and B of Fig. 2. The state of polarization of each slice is manually tuned by mean of polarization controllers placed after the WSS. To combine the 2 slices with the correct phase offset we engineer a small spectral overlap region between them (8-GHz width). The interference in the overlap region of the spectrum produces an error signal $U_{err}$, which is approximately proportional to the phase error $\Delta\varphi$ between the slices. This error signal is exploited by the signal-combining element (SCE) which implements a feedback-control loop to stabilize $\Delta\varphi$ [12]. The 300 GBd signal is then sent to a polarization division multiplexing emulator (PDME) based on a split and delay architecture. The relative delay between both polarizations is ~84 ns. The channel under test is then combined with dummy channels, each carved out of spectrally white amplified spontaneous emission (ASE) noise using a programmable WaveShaper, emulating a channel spacing of 306.25 GHz. The resulting WDM comb, shown in Fig. 2(c), is injected into the recirculating loop. The submarine recirculating loop consists of 11 spans of 55-km Corning EX3000 fibers, with 0.157-dB/km loss coefficient, +20.5-ps/(nm.km) dispersion at 1550 nm, +0.06-ps/(nm$^2$.km) dispersion slope at 1550 nm, and 150-µm$^2$ effective area. Each loop thus corresponds to transmission over 605-km. Fiber attenuation is compensated at the end of each span of the loop by a C-band EDFA followed by a gain flattening filter (GFF). Channel power equalization across the WDM comb is performed after the last span of the loop, as shown in Fig. 2. Finally, a loop synchronous polarization scrambler (PS) enables randomly distributing the state of polarization of the signals over the Poincaré sphere after each loop round trip. At the receiver, we employ a 2-slice dual-polarization OAWM receiver, which uses two phase-locked local oscillator (LO) tones at frequencies $f_{Rx,1} \approx f_{Tx,1}$ and $f_{Rx,2} \approx f_{Tx,2}$ for coherent detection. The phase-locked tones are spaced by 144 GHz and selected from a receiver frequency comb (Rx comb) generated by modulating a CW laser tone with a 36 GHz sin wave, as shown in Fig. 2. To get an identical state of polarization for each slice before entering the coherent receivers, we again use manual polarization controllers, as shown in Fig. 2. It must be noted that future development of integrated OAWG and OAWM through photonic integrated circuits [13] will suppress the need for manual tuning of the state of polarization (both at Tx and Rx). We simultaneously capture two spectral slices for each polarization by using eight channels of two synchronized Keysight UXR series oscilloscopes. Each recorded waveform consists of 8×2 million samples which are processed offline. We first compensate for the phase offset occurring at the receiver between the 2 slices. To do so, we use the small spectral overlap region between the traces recorded by the 2 oscilloscopes (each having an electricalbandwidth greater than half the receiver comb FSR). The phase offset is obtained by computing arg($\underline{S}_1(f) \times \underline{S}_2(f)^*$) with $\underline{S}_1(f)$ and $\underline{S}_2(f)$ being the FFT of each slice and * denoting the complex conjugation. In the spectral overlap region, $\underline{S}_1(f)$ and $\underline{S}_2(f)$ carry identical information such that the argument of $\underline{S}_1(f) \times \underline{S}_2(f)^*$ is the phase offset to compensate for. Following, the standard coherent DSP suite consists of chromatic dispersion compensation, complex 2x2 MIMO adaptive equalization, decimation at 1 sample per symbol, frequency offset and carrier phase recovery, and least-mean square equalizer to mitigate transmitter I/Q imbalances. Adaptive equalization is done using periodically distributed QPSK pilots with a rate of 2.05% [14]. No digital backpropagation for nonlinear effects compensation is applied. Using the demodulated signal, we independently measure the signal-to-noise ratio (SNR), the normalized generalized mutual information (NGMI), and the maximum achievable net bit rate by applying a flexible FEC rate based on a family of SC LDPC codes with rates varying between 0.5 and 0.9 to achieve error-free decoding [14].

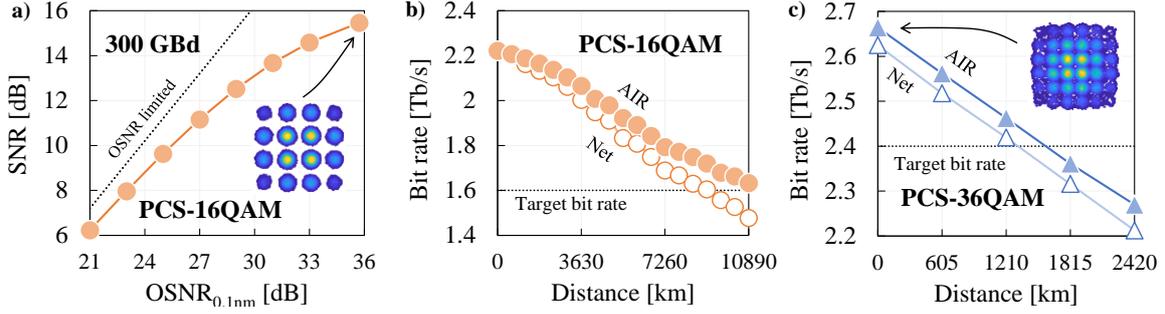

**Fig. 3:** Experimental results at 300 GBd. a) Back-to-back sensitivity to noise; b-c) Bit rate versus distance for PCS-16QAM and PCS-36QAM.

**Transmission results**
We first perform back-to-back experiments to assess the performance of the 300 GBd dual polarization optically sliced transceiver. The signal entropy is set at 3.8 bits/symbol/polarization for PCS-16QAM. The Fig. 3(a) shows the SNR as a function of the optical SNR (OSNR) measured in 0.1 nm bandwidth. This graph shows that the SNR saturates at ~15 dB for high OSNR values, owing for imperfections and limited bandwidth of each coherent driver modulator. An exemplary constellation diagram for the best SNR value is shown in Fig. 3(a). We then perform submarine transmission experiments. The output power of line EDFAs is set to 17 dBm, corresponding to the optimum launch power after 6655 km transmission distance [15]. Figure 3(b) shows the measured achievable information rate (AIR) and net bit rate after FEC decoding versus transmission distance. For each transmission distance, among the family of SC-LDPC codes, we select the one having the highest code rate $r_c$ enabling error free decoding and the net bit rate is obtained from the relation $Net\ bit\ rate = 2 \cdot R \cdot (1 - R_p) \cdot (H - (1 - r_c) \cdot m)$ with $R$ the symbol rate, $R_p$ the pilot overhead correction, $H$ the PCS signal entropy and $m$ the native constellation cardinality (4 and 6 bits/symbol/polarization for PCS-16QAM and PCS-36QAM, respectively). The AIR is obtained by replacing $r_c$ by NGMI in the above equation. Fig. 2(b) shows that an AIR of 1.6 Tb/s is achieved at up to 10890 km while 1.6 Tb/s net bit rate is obtained at 9075 km with $r_c$=0.74, resulting in a net spectral efficiency of 5.22 bits/s/Hz. With respect to previous achievements, this constitutes a 10x improvement in transmission distance for aggregated capacity per carrier of 1.6 Tb/s. Finally, we perform the transmission of PCS-36QAM with an entropy of 5 bits/symbol/polarization. The results are reported in Fig. 3(c). The net data rate of 2.4 Tb/s is achieved at 1210 km with a code rate $r_c$= 0.85 and a net spectral efficiency of 7.8 bits/s/Hz. This constitutes the first demonstration of long-haul signal-carrier 2.4 Tb/s transmission.

**Conclusion**
We demonstrate the first transatlantic transmission of 1.6 Tb/s single carrier using 300 GBd dual polarization PCS-16QAM signal based on 2-slice optical arbitrary waveform generation and measurement. We also achieve the longest transmission distance for single-carrier 2.4 Tb/s over 1210 km. These results represent a first step towards next generation of long-haul coherent systems running at 1.6 Tb/s and 2.4 Tb/s while highlighting the benefit of optical multiplexing approaches to achieve ultra-high symbol rates.

**Funding Acknowledgements**
The authors acknowledge Keysight Technologies for supporting this work by providing one UXR sampling scope. This work was supported by the EIC Transition projects CombTools (# 101136978) and HDLN (# 101113260), by the HORIZON-ERC-2023-POC Teragear (# 101123567), by the ERC Consolidator Grant TeraSHAPE (# 773248), by the DFG projects PACE (# 403188360) and GOSPEL (# 403187440), by the DFG Collaborative Research Centers (CRC) WavePhenomena (SFB 1173, # 258734477) and HyPERION (SFB 1527, # 454252029), by the BMBF project Open6GHub (# 16KISK010), by the Alfried Krupp von Bohlen und Halbach Foundation, by the MaxPlanck School of Photonics (MPSP), and by the Karlsruhe School of Optics & Photonics (KSOP).